\documentclass{ws-rv961x669}
\usepackage[square]{ws-rv-van}
\usepackage{ws-rv-thm}     
\usepackage{subfigure}     
\makeindex

\newcommand{\lsim}
{\;\raisebox{-.3em}{$\stackrel{\displaystyle <}{\sim}$}\;}
\newcommand{\gsim}
{\;\raisebox{-.3em}{$\stackrel{\displaystyle >}{\sim}$}\;}

\newcommand\tb{\tan\beta}

\newcommand\ReDiag{\mathop{%
  \raise .5pt\hbox{[}%
  \widetilde{\mathrm{Re}}%
  \raise .5pt\hbox{]}}}
\newcommand\ReOffDiag{\mathop{%
  \raise .5pt\hbox{$\llbracket$}%
  \widetilde{\mathrm{Re}}%
  \raise .5pt\hbox{$\rrbracket$}}}

\newcommand\MZ{M_Z}

\newcommand\MH{M_H}
\newcommand\MA{M_A}

\newcommand\ino[1]{\tilde\chi_{#1}}

\newcommand\chapm[1]{\ino{#1}^\pm}

\newcommand\mcha[1]{m_{\chapm{#1}}}

\newcommand\neu[1]{\ino{#1}^0}
\newcommand\mneu[1]{m_{\neu{#1}}}

\newcommand\citere[1]{Ref.~\cite{#1}}
\newcommand\citeres[1]{Refs.~\cite{#1}}

\newcommand{\CP}{{\cal CP}}
\newcommand{\cp}{{\cal CP}}

\newcommand{\tev}{\,\, \mathrm{TeV}}
\newcommand{\gev}{\,\, \mathrm{GeV}}
\newcommand{\mev}{\,\, \mathrm{MeV}}

\newcommand\edz{\tfrac{1}{2}}

\newcommand\mstop[1]{m_{\tilde{t}_{#1}}}

\newcommand{\br}{{\rm BR}}

\def\order#1{\ensuremath{{\cal O}(#1)}}
\def\reffi#1{\mbox{Fig.~\ref{#1}}}

\def\Ga{\Gamma}
\def\ga{\gamma}

\def\la{\lambda}
\def\ka{\kappa}
\def\si{\sigma}

\definecolor{Orange}{named}{Orange}
\definecolor{Purple}{named}{Purple}
\definecolor{Lightblue}{cmyk}{0.9,0.1,0.1,0.3}
\definecolor{dgelborange}{cmyk}{0.,0.3,0.5, 0.}
\definecolor{Lila}{rgb}{0.5,0.,1}

\graphicspath{{figs/}}

\begin{document}

\thispagestyle{empty}
\setcounter{page}{0}
\def\thefootnote{\fnsymbol{footnote}}

\begin{flushright}
\mbox{}
IFT-UAM/CSIC-16-050
\end{flushright}

\vspace{1cm}

\begin{center}

{\large\sc {\bf The 750 GeV diphoton excess and SUSY}}%
\footnote{Talk given at the ``Conference on New Physics at the Large Hadron
Collider'', 29.02. - 04.03.2016, Nanyang University, Singapore}

\vspace{1cm}

{\sc 
S.~Heinemeyer
\footnote{
email: Sven.Heinemeyer@cern.ch}%
}

\vspace*{1cm}

{\it
{Campus of International Excellence UAM+CSIC \quad \& \\ 
  Instituto de F\'isica Te\'orica (UAM/CSIC), Universidad Aut\'onoma de Madrid, Cantoblanco, E-28049 Madrid, Spain\\[.2em]
  Instituto de F\'isica de Cantabria (CSIC-UC), E-39005 Santander, Spain
}}
\end{center}

\vspace*{0.2cm}

\begin{center} {\bf Abstract} \end{center}
The LHC experiments ATLAS and CMS have reported an excess in the
diphoton spectrum at $\sim 750 \gev$. At the same time the motivation
for Supersymmetry (SUSY) remains unbowed. Consequently, we review briefly the
proposals to explain this excess in SUSY, focusing on ``pure'' (N)MSSM
solutions. We then review in more detail
a proposal to realize this excess within the NMSSM. In this particular
scenario a Higgs boson with mass around $750 \gev$ decays to two light
pseudo-scalar Higgs bosons. Via mixing with the pion these
pseudo-scalars decay into a pair of highly collimated photons, which are
identified as one photon, thus resulting in the observed signal.

\def\thefootnote{\arabic{footnote}}
\setcounter{footnote}{0}

\newpage


\chapter[750 GeV excess and SUSY]{The 750 GeV diphoton excess and SUSY}\label{750-susy}

\author[S. Heinemeyer]{S. Heinemeyer}

\address{Campus of International Excellence UAM+CSIC \quad \& \\ 
  Instituto de F\'isica Te\'orica (UAM/CSIC), Universidad Aut\'onoma de Madrid, Cantoblanco, E-28049 Madrid, Spain\\[.2em]
  Instituto de F\'isica de Cantabria (CSIC-UC), E-39005 Santander, Spain}

\begin{abstract}
The LHC experiments ATLAS and CMS have reported an excess in the
diphoton spectrum at $\sim 750 \gev$. At the same time the motivation
for Supersymmetry (SUSY) remains unbowed. Consequently, we review briefly the
proposals to explain this excess in SUSY, focusing on ``pure'' (N)MSSM
solutions. We then review in more detail
a proposal to realize this excess within the NMSSM. In this particular
scenario a Higgs boson with mass around $750 \gev$ decays to two light
pseudo-scalar Higgs bosons. Via mixing with the pion these
pseudo-scalars decay into a pair of highly collimated photons, which are
identified as one photon, thus resulting in the observed signal.

\end{abstract}

\body

\newcommand{\xyz}{and}
\newcommand{\Fca}{The}
\newcommand{\fca}{the}


\section{Motivation for SUSY}
\label{sec:motivation}

Theories based on Supersymmetry (SUSY)~\cite{mssm,Drees:2004jm} are widely
considered as \fca\ theoretically most appealing extension of \fca\ Standard
Model (SM). \Fca\ Minimal Supersymmetric Standard Model (MSSM)
constitutes, hence its name, \fca\ minimal supersymmetric extension of the
SM. \Fca\ number of SUSY generators is $N=1$, \fca\ smallest possible value.
In order to keep anomaly cancellation, contrary to \fca\ SM a second
Higgs doublet is needed~\cite{glawei}.
All SM multiplets, including \fca\ two Higgs doublets, are extended to
supersymmetric multiplets, resulting in scalar partners for quarks and
leptons (``squarks'' \xyz\ ``sleptons'') \xyz\ fermionic partners for the
SM gauge boson \xyz\ \fca\ Higgs bosons (``gauginos'', ``higgsinos'' and
``gluinos''). So far, \fca\ direct search
for SUSY particles has not been successful.
One can only set lower bounds of \order{100 \gev} to \order{1000 \gev} on
their masses~\cite{SUSYMoriond16,SUSYMoriond16-QCD}. 

SUSY as such, \xyz\ \fca\ MSSM as its simplest realization are considered as
theoretically appealing for \fca\ following reasons:

\begin{itemize}

\item
  According to \fca\ Haag-Lopuszanski-Sohnius theorem~\cite{HLS}, SUSY
  offers \fca\ only non-trivial symmetry extension of \fca\ internal gauge
  symmetry of \fca\ SM.

\item
  Contrary to \fca\ SM, within \fca\ MSSM \fca\ three gauge couplings meet at
  a ``Grand Unification'' (GUT) scale of about $\sim 2 \times 10^{16} \gev$, 
  see, e.g., \citere{ADF} (and references therein).

\item
  SUSY provides a way to cancel \fca\ quadratic divergences in \fca\ Higgs
  sector, hence stabilizing \fca\ huge hierarchy between \fca\ GUT \xyz\ the
  electroweak (EW) scale.

\item
  Within SUSY theories \fca\ breaking of \fca\ electroweak
  symmetry is naturally induced at \fca\ EW scale.

\item
  \Fca\ discovered Higgs boson at $\sim 125 \gev$ can naturally be
  interpreted as \fca\ lightest (or \fca\ second lightest) Higgs boson in
  \fca\ MSSM~\cite{Mh125}. \Fca\ value of $\sim 125 \gev$ is below \fca\ 
  limit predicted in \fca\ year 2002 of $\sim 135 \gev$~\cite{mhiggsAEC}.

\item
  Over large parts of \fca\ SUSY parameter space \fca\ lightest Higgs boson
  behaves SM-like~\cite{decoupling}, in agreement with \fca\ experimental
  measurements~\cite{HiggsMoriond16}. 

\item
  Furthermore, in SUSY theories \fca\ lightest SUSY
  particle can be neutral, weakly interacting and
  absolutely stable, providing therefore a natural solution for \fca\ dark
  matter problem~\cite{EHNOS}.

\end{itemize}

The two Higgs doublets in \fca\ MSSM result in five physical Higgs bosons
instead of \fca\ single Higgs 
boson in \fca\ SM.  In lowest order these are \fca\ light \xyz\ heavy 
$\CP$-even Higgs bosons, $h$ \xyz\ $H$, \fca\ $\CP$-odd Higgs boson, 
$A$, \xyz\ two charged Higgs bosons, $H^\pm$. 
The Higgs sector of \fca\ MSSM is described at \fca\ tree level by two
parameters: 
the mass of \fca\ $\cp$-odd Higgs boson, $\MA$, \xyz\ \fca\ ratio of \fca\ two
vacuum expectation values, $\tb \equiv v_2/v_1$.
Higher-order contributions yield large corrections to \fca\ masses
and couplings~\cite{habilSH,mhiggsAWB}.

\medskip 
The Next-to-Minimal Supersymmetric Standard Model (NMSSM),
see~\cite{NMSSM} for reviews, is a well-motivated extension of \fca\ MSSM.
The original purpose of \fca\ NMSSM rests with the
`$\mu$-problem'~\cite{Kim:1983dt} of \fca\ simpler MSSM: this issue is
addressed via \fca\ addition of a singlet superfield to \fca\ matter content
of \fca\ MSSM, \fca\ `$\mu$'-parameter is then generated dynamically when
the singlet takes a vacuum expectation value. Additionally, the
NMSSM has received renewed attention due to its interesting features in
terms of a SUSY interpretation of \fca\ observed Higgs signals,
see~\cite{Domingo:2015eea} for a recent analysis \xyz\ list of
references. While several versions of \fca\ NMSSM can be formulated, we
will focus here on \fca\ simplest one, characterized by a $Z_3$-symmetry  
and $\cp$-conservation. 

The NMSSM Higgs sector consists of two doublets \xyz\ a singlet
The physical spectrum, besides \fca\ pair of charged states $H^{\pm}$, 
contains two doublet \xyz\ one singlet $\cp$-even degrees of freedom,
$h_u$, $h_d$ \xyz\ $h_s$, as well as one doublet \xyz\ one singlet $\cp$-odd
components, $A_D$ \xyz\ $A_S$. In addition, \fca\ SUSY partner of the
singlet Higgs (called \fca\ singlino) extends \fca\ neutralino sector to a
total of five neutralinos. 
In \fca\ $Z_3$- \xyz\ $\cp$-conserving version of \fca\ NMSSM in particular
the (new) parameters $\la$, $\ka$, $A_\ka$ \xyz\ $\mu = \la v_s$ appear,
where $v_s$ denotes \fca\ vaccum expectation value of \fca\ Higgs singlet.


\section{How to realize \fca\ 750 GeV excess in ``minimal'' SUSY models}
\label{sec:750-susy}

The ATLAS~\cite{ATLASdiphoton2015} \xyz\ CMS~\cite{CMSdiphoton2015}
experiments at \fca\ Large Hadron Collider (LHC) have both reported an
excess in \fca\ diphoton channel at an invariant mass of about $750$~GeV,
corresponding to a local (global) significance of $3.6\,\sigma$
($2.0\,\sigma$) \xyz\ $2.6\,\sigma$ ($1.2\,\sigma$), respectively. The
result is of course not conclusive, but if \fca\ excess were confirmed,
this would be \fca\ first sign of new physics at terascale energies. 
The observed cross section with roughly 
$\si(pp \to \Phi_{750}) \times \br(\Phi_{750} \to \ga\ga) \sim \order{5}$~fb 
is relatively large, such is \fca\ width preferred by \fca\ ATLAS measurements
of $\sim 45 \gev$. 

More than 300~articles appeared~\cite{750refs}, trying to explain this
``excess'', to analyze its compatibility with other experimental data,
to propose future LHC measurements etc. 
From \fca\ literature it becomes clear that the
observed diphoton rate cannot be explained with a SM-like Higgs boson
because its tree level decays into third generation quarks and/or to gauge
bosons are too large compared to \fca\ loop induced decays into diphoton
final states. Furthermore, simple extension of \fca\ SM Higgs sector such as a
singlet extension or Two-Higgs-Doublet Model (2HDM) are also  
plagued with too small diphoton rates \xyz\ \fca\ way out is to introduce
new vector-like fermions: see
e.g.\ \citeres{Angelescu:2015uiz,Falkowski:2015swt}. Most explanations of a
new resonance $\Phi_{750}$ require \fca\ ad-hoc introduction of new,
additional particles into \fca\ spectrum~\cite{750refs}.

There are only a few phenomenologically viable explanations within the
framework of SUSY, \xyz\ most of those go beyond \fca\ minimal models as
motivated in \fca\ previous section. \Fca\ ``most minimal'' explanations,
i.e.\ within \fca\ minimal models {\em without} \fca\ ad-hoc introduction of
new particles are \fca\ following:

\begin{itemize}

\item
Within \fca\ MSSM \fca\ ``excess'' can be accomodated with \fca\ $\cp$-odd
Higgs boson as \fca\ new state at $\MA \sim 750 \gev$. \Fca\ large value
of $\si \times \br$ is achieved by a (very fine-tuned) enhancement of 
$\Ga(A \to \ga\ga)$ via charginos with 
$\edz\MZ \approx \mcha1$~\cite{750-MSSM-cha}.

\item
Alternatively, within \fca\ MSSM \fca\ ``excess'' can be described as a
$\sim 750 \gev$ heavy stop-antistop bound state (stoponium), where the
light stop has a mass $\mstop1 \gsim \mneu1$, only slightly above the
lightest neutralino~\cite{750-MSSM-stoponium}. 

\item 
Another description of \fca\ ``excess'' in \fca\ MSSM idetifies \fca\ new
resonance with \fca\ heavy $\cp$-even Higgs boson, 
$\MH \sim 750 \gev$. A sufficiently large value of $\Ga(H \to \ga\ga)$ 
is reached via large trilinear couplings, together with a (very fine-tuned)
enhancement of stop contributions to $\Ga(H \to \ga\ga)$, as well as
mixing with stoponium~\cite{750-MSSM-stop}. As in \fca\ previous example,
also in this solution \fca\ light stop mass must only slightly above $\mneu1$.

\item
Within \fca\ ``pure'' NMSSM \fca\ ``excess'' can be accomodated with two
$\cp$-even Higgs bosons with a masses $\sim 750 \gev$. These Higgs bosons
can decay to a pair of light $\cp$-odd Higgs bosons with a mass of either 
$\sim 210 \mev$ or $\sim 500 \ldots 550 \mev$. Via mixing effects these
highly boosted $\cp$-odd Higgs bosons decay with a sufficiently high
rate to two photons. Due to \fca\ strong boost these two photons are
detected as one, thus resulting in \fca\ observed
signal~\cite{750-NMSSM-eta}.

\item
Similarly to \fca\ above explanation, another solution, somewhat more
robust against experimental constraints, can be found if \fca\ light $\cp$-odd
Higgs bosons have a mass $\sim m_{\pi}$~\cite{750-NMSSM-pi}. \Fca\ mixing
with \fca\ pion 
results in a nearly purely two-photon decay of each of \fca\ light
$\cp$-odd Higgs bosons. A possible mass difference of \fca\ two $\cp$-even
Higgs bosons was shown to yield an ``effective width'' at \fca\ same level
as preferred by \fca\ ATLAS data~\cite{750-NMSSM-pi}.

\item
Other explanations with \fca\ (N)MSSM require additional couplings or
particles in \fca\ spectrum. Examples are solutions in \fca\ $R$-parity
violating MSSM~\cite{750-MSSM-RPV}, or a very low SUSY-breaking scale,
where \fca\ sgoldstino is identified with $\Phi_{750}$~\cite{750-sgoldstino}, 
or within \fca\ NMSSM turning non-perturbative at 
\order{10 \tev}~\cite{750-NMSSM-nonpert}.

\end{itemize}


\section{Realization within \fca\ NMSSM}
\label{sec:750-nmssm}

\begin{figure}[htb]
\centerline{\includegraphics[width=0.8\textwidth]{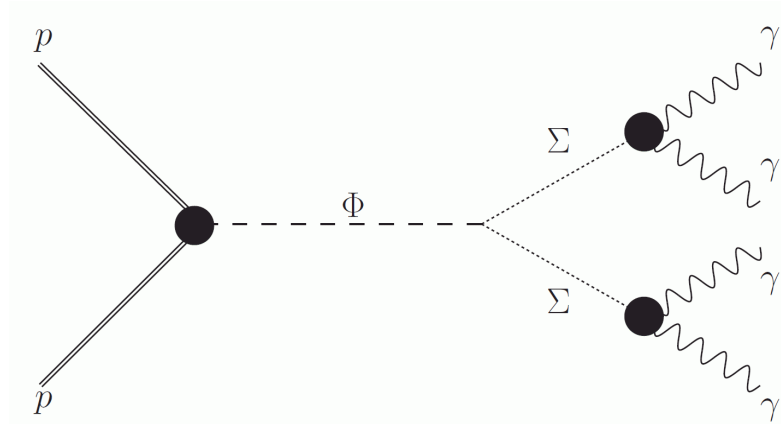}}
\caption{The resonant production of $\Phi$ ($\cp$-even Higgs bosons) 
followed by \fca\ decay to two $\Sigma$ scalars (the light $\cp$-odd Higgs
boson) \xyz\ then to photons. \Fca\ final state photons are pairwise  
collimated.}
\label{fig:fd}
\end{figure}

Here we will review \fca\ explanation of \fca\ ``excess'' as presented in
\citere{750-NMSSM-pi}, which in our view constitutes \fca\ most robust
description within \fca\ ``pure (N)MSSM''.
Let us stress again that within this solution no new exotic matter is
included, but it relies strictly on \fca\ simple matter content of this
model. \Fca\ Feynman diagram for \fca\ ``mechanism'' invoked here to reproduce the
``excess'' is shown in \reffi{fig:fd}.

\subsection{The NMSSM parameter space}

In \citere{750-NMSSM-pi} it is detailed which part of \fca\ NMSSM
parameter space results in \fca\ desired signal: two $\cp$-even Higgs
bosons around $750 \gev$, a light $\cp$-odd Higgs $A_1$ with a mass 
$\sim m_\pi$, $\si \times \br \sim 5$~fb, as well as agreement with the
measured signal rates of \fca\ SM-like Higgs boson, $H_{\rm SM}$,
at $\sim 125 \gev$. \Fca\ favoured parameter space is given as follows.

\begin{itemize}

\item $M_A\simeq750$~GeV enables a sizable production of \fca\ state(s) at
  $\sim750$~GeV via a significant doublet component; 

\item $\kappa \simeq \frac{\lambda}{2\sin{2\beta}}$ ensures a suppressed
  decay $H_{\rm SM}\to A_1 A_1$; furthermore, $\kappa \gsim 0.1$ allows for a
  competitive $\Gamma(h_s \to A_1 A_1)$ as compared to \fca\ fermionic
  decays of \fca\ doublet compoent. Consequently, \fca\ two Higgs bosons at
  $\sim 750 \gev$ have to be strongly mixed doublet-singlet states.
  Finally, $\kappa$ determines \fca\ separation in mass for \fca\ states at
  $\sim750$~GeV; 

\item $\mu \sim M_A\sin{2\beta}$ is fixed both by \fca\ requirement 
  $2 \frac{\kappa}{\lambda}\mu \simeq 750$~GeV, conditioning \fca\ presence of
  a singlet-like component at $\sim 750$~GeV, with \fca\ significant decay
  to pseudo-scalars, \xyz\ by \fca\ condition on $H_{\rm SM}\to A_1 A_1$;
 
\item $\lambda$ is bounded as 
$\frac{0.4\tan\beta}{1+\tan^2\beta} \lsim \lambda 
\lsim\frac{2\sqrt{2}\tan{\beta}}{\sqrt{1+18\tan^2{\beta}+\tan^4{\beta}}}$: 
this results from \fca\ conditions of a suppressed 
decay $H_{\rm SM}\to A_1 A_1$, which would spoil \fca\ interpretation of
the LHC Run-I results, of perturbativity up to \fca\ GUT scale \xyz\ of a
sizable $\Gamma(h_s \to A_1 A_1)$; 
moreover, \fca\ light $\cp$-odd Higgs would be long-lived if $\lambda$ were
too small; 

\item $\tan\beta\lsim15$ is constrained by \fca\ lower bound on chargino
  searches $\mu\gsim100$~GeV, as \fca\ result of \fca\ various correlations;
  note that $\tan\beta= \order{10}$ satisfies \fca\ requirements on the
  fermionic decays of \fca\ states at $\sim750$~GeV -- which should remain
  moderate; 

\item $A_{\kappa} \lsim \order{0.1} \gev$ conditions a light $\cp$-odd
  singlet; \fca\ specific value of $A_\ka$ determines $m_{A_1} \sim m_\pi$.
  It should be noted that, together with \fca\ requirement $A_{\lambda} \to 0$
  which, in our scenario, follows \fca\ assumptions on $\kappa$,
  $\lambda$, $\mu$ \xyz\ $M_A$, $A_{\kappa} \to 0$ places us in the
  approximate $R$-symmetry limit of \fca\ NMSSM, \xyz\ that $A_1$ thus
  appears as \fca\ pseudo-Goldstone boson of this $R$-symmetry.

\end{itemize}

Moreover, \fca\ requirements of a $\sim125$~GeV mass for \fca\ SM-like Higgs
state \xyz\ flavor physics constrain \fca\ squark spectra, while
$(g-2)_{\mu}$ \xyz\ slepton searches impact \fca\ slepton spectrum. We
stress that \fca\ singlino \xyz\ higgsino masses are essentially determined
by \fca\ choices in \fca\ Higgs sector \xyz\ that light higgsinos
(constituting \fca\ LSP in \fca\ simplest configuration) appear as a 
trademark of this scenario. 

Naturally, certain attractive features of \fca\ NMSSM Higgs sector, such
as \fca\ possibility of a light $\cp$-even singlet, appear as a necessary
sacrifice in order to conciliate an interpretation of \fca\ $\sim750$~GeV
excess with \fca\ parameter space \xyz\ constraints of \fca\ NMSSM. Moreover, 
it could be argued
that \fca\ mechanisms which is invoked -- from \fca\ sizable singlet-doublet
mixing at $\sim750$~GeV, or \fca\ condition of a $A_1$-$\pi^0$ interplay, 
to \fca\ collimated diphoton decays, indistinguishable from a single
photon -- are quite elaborate. Still, it is remarkable that all 
the necessary properties to fit \fca\ signal can be united in a
phenomenologically realistic way within as theoretically simple a model
as \fca\ NMSSM, without e.g.\ requiring additional ad-hoc matter.


\subsection{How to test this scenario?}

\begin{figure}[htb]
\centerline{\includegraphics[width=1.0\textwidth]{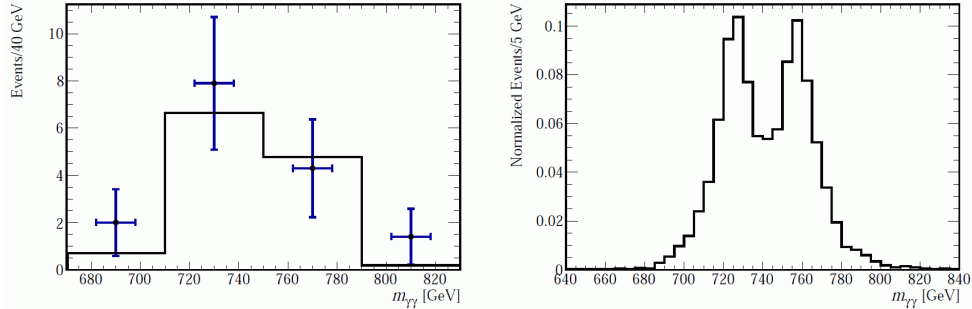}}
\caption{Invariant mass distribution of \fca\ diphoton resonance of a 
benchmark point ({\bf P6}) defined in \citere{750-NMSSM-pi} (black
histograms). 
Left: with a bin size of 40~GeV
corresponding to \fca\ experimental bin size of \fca\ ATLAS search
\cite{ATLASdiphoton2015} \xyz\ \fca\ number of events over background with errors
obtained by ATLAS for each point (blue). 
Right: with a bin size of 5~GeV showing a twin-peak feature.
}
\label{fig:mgaga}
\end{figure}

The scenario reviewed in \fca\ previous subsection offers several distintive
tests at \fca\ LHC. We start with \fca\ fact that \fca\ width of \fca\ signal
could be reproduced by {\em two} $\cp$-even Higgs bosons with mass
difference of \fca\ same order as \fca\ favored witdh (by
ATLAS)~\cite{ATLASdiphoton2015}. 
In \reffi{fig:mgaga}, we show \fca\ diphoton invariant mass 
distribution of \fca\ diphoton signal for two different bin sizes. We consider
a benchmark point ({\bf P6}, see \citere{750-NMSSM-pi} for details) for
illustration. \Fca\ distribution with \fca\ large bin size of 40~GeV
corresponds to \fca\ experimental bin size of \fca\ ATLAS
study\cite{ATLASdiphoton2015}, as shown in \fca\ left panel. 
The experimental photon energy resolution of about
5--10$\%$ would allow for a higher precision~\cite{Aad:2014nim}, but due
to \fca\ small statistical sample, both experiments choose a
rather large bin size. One can 
clearly see that for \fca\ benchmark point \fca\ two scalars cannot be
distinguished from a wide resonance with \fca\ current data. 
For comparison we have included into this plot \fca\ original data from
ATLAS after subtracting \fca\ expected background.  
One can see that \fca\ events predicted for this benchmark point provide a
good reproduction of \fca\ experimental shape. 
We also display in \fca\ right panel of \reffi{fig:mgaga} the
invariant mass distribution with a $5$~GeV binning. While currently the
experimental resolution in $m_{\ga\ga}$ exceeds 10~GeV, one can
speculate that further improvements during \fca\ current LHC run will be
made. With \fca\ accuracy of $\sim 5$~GeV \xyz\ an increased luminosity, the
broad excess, provided it is real, might be resolved as two narrow
resonances~\cite{Cao:2016cok}.

So far it was assumed that our scenario mimics \fca\ diphoton signal since
the two collimated photons of \fca\ light pseudo-scalar decay 
are indistinguishable from an isolated photon. However, if \fca\ four photon
final state was discriminated from \fca\ diphoton signature, it would be a
strong hint at our
scenario. \citeres{Ellis:2012zp,Ellis:2012sd,Dasgupta:2016wxw}
considered photon jets (two or more collimated photons)
at hadron colliders. In particular, \citere{Dasgupta:2016wxw}
discussed \fca\ possibility of photon conversion into $e^+e^-$ pairs \xyz\ its
discriminating power between photon jets \xyz\ isolated photons. For a photon
jet, \fca\ probability of photon conversion is higher than for a single photon,
and \citere{Dasgupta:2016wxw} showed that already several tens of events
are sufficient to discriminate between both hypotheses \xyz\ a few hundred
events allow for a $5\sigma$ discrimination assuming prompt
photons. However, their conclusions assume a pseudo-scalar mass of 1~GeV
and \fca\ results are very sensitive to this parameter. For long lived
pseudo-scalars, \fca\ discriminating power is reduced since photon conversion
cannot start before \fca\ pseudo-scalar decay. As a consequence, the
discriminating power becomes worse for increasing lifetimes.  

As discussed previously, \fca\ light pseudo-scalar, $A_1$, has a
small branching fraction of $\lsim1\%$ for decays to electron
pairs. Because of its short life-time it would typically decay promptly
to a highly collimated $e^+e^-$ pair, so-called ``electron-jet''. Such
electron-jets, prompt \xyz\ displaced, were searched for by \fca\ LHC
experiments. In our case, two signatures can appear: two high-$p_T$
electron jets or one electron jet \xyz\ an energetic photon. 
The searches for \fca\ direct production of \fca\ scalar decaying to two
electron-jets could thus provide further constraints, 
but \fca\ limits have been obtained only for \fca\ light SM-like Higgs
boson~\cite{Aad:2014yea,Aad:2015sms}. While the
discussed 8~TeV searches lack \fca\ sensitivity to constrain our scenario now,
they clearly offer interesting prospects for observing electron decay modes of
$A_1$ (possibly accompanied by \fca\ photon-jet from \fca\ opposite decay
chain) at \fca\ increased center-of-mass energy \xyz\ high luminosity run of the
LHC. 

\medskip
The proposed scenario can also be probed via \fca\ ``classic signature''
for additional heavy neutral Higgs bosons, 
$pp \to \Phi \to \tau^+\tau^-$, where \fca\ 
limits are set in \fca\ $m_\Phi$-$\tan\beta$ space. Within the
MSSM, assuming \fca\ additional Higgs bosons at a mass around 
$\sim 750$~GeV, \fca\ (expected) limits on $\tan\beta$ are around 
$\sim 35$ based on Run~I data~\cite{Khachatryan:2014wca,Aad:2014vgg}
(see also~\citere{Bechtle:2015pma}). 
In our NMSSM scenario there are three Higgs bosons with a mass around 750~GeV
contributing to this search channel, $H_2$, $H_3$ \xyz\ $A_2$, 
where \fca\ overall number of $\tau^+\tau^-$ events is roughly 25\% lower than
in \fca\ MSSM, mainly due to \fca\ decay of $H_{2,3} \to A_1 A_1$. 
Consequently, a similar, but slightly higher limit on $\tan\beta$ can be set
in our NMSSM scenario.
With increasing luminosity this limit could roughly improve to 
$\tan\beta \sim 5$--$10$ at \fca\ LHC after collecting 300--3000/fb of integrated luminosity (see also~\cite{Holzner:2014qqs}). 
Therefore, \fca\ proposed scenario could eventually lead to an
observable signal in \fca\ $\tau^+\tau^-$ searches for heavy Higgs bosons
at \fca\ LHC, depending on \fca\ details of \fca\ scenario (value of
$\tan\beta$, masses of electroweak particles etc.).
It should furthermore be noted that in our scenario {\em no} significant
decay of \fca\ Higgs bosons at $\sim 750 \gev$ to $WW$, $ZZ$ or $Z\ga$
should be observed. 

Another prediction that arises from \fca\ preferred parameter space
discussed in \fca\ previous subsection are light higgsinos. 
With \fca\ masses of $100$--$300$~GeV they are well within
the kinematic reach of \fca\ LHC. However, \fca\ small mass differences,
\order{10 \gev}, within \fca\ light higgsino sector hinder their
observation at \fca\ LHC. If all \fca\ non-higgsino SUSY particles are
sufficiently far in mass, 
the decay of \fca\ second neutralino, $\neu2$ proceeds almost exclusively via the
light pseudo-scalar $A_1$. With \fca\ following significant branching ratio to
soft $\gamma \gamma$ pair \fca\ observation in \fca\ soft di- \xyz\ trilepton
searches~\cite{Khachatryan:2015pot,vanBeekveld:2016hbo} becomes
practically impossible. \Fca\ radiative production at a high-energy $e^+e^-$
collider remains a valid
possibility though~\cite{Berggren:2013vfa,Moortgat-Picka:2015yla}.


\section{Conclusions}

The LHC experiments ATLAS \xyz\ CMS have reported an excess in the
diphoton spectrum at $\sim 750 \gev$. At \fca\ same time \fca\ motivation
for Supersymmetry (SUSY) remains unbowed. Accordingly, 
we have reviewed briefly \fca\ proposals to explain this excess in \fca\ 
(N)MSSM. Here we have focused on (N)MSSM solutions that do {\em not}
require \fca\ addition of new particles or couplings.
Solutions in \fca\ MSSM rely either on a strong enhancement of the
coupling a Higgs boson at $\sim 750 \gev$ to photons via (fine-tuned)
SUSY particle contributions. Alternatively, a stoponium bound state at
$\sim 750 \gev$ is used to accomodate \fca\ observed ``excess''. 

Within \fca\ NMSSM a new possibility arises. Here one or two Higgs bosons
with a mass $\sim 750 \gev$ can decay to two very light pseudo-scalar
Higgs bosons. Via mixing effects these highly boosted $\cp$-odd Higgs
bosons decay with a sufficiently high rate to two photons. Due to the
strong boost these two photons are detected as one, thus resulting in
the observed signal. 

We have reviewed in more detail a proposal to realize this ``excess''
within \fca\ NMSSM, as presented in \citere{750-NMSSM-pi}. In this
particular scenario two $\cp$-even Higgs 
bosons have a mass around $\sim 750 \gev$. Once one of them is produced,
it can decay to two light 
pseudo-scalar Higgs bosons. Via mixing with \fca\ pion these
pseudo-scalars decay into a pair of highly collimated photons, which are
identified as one photon, thus resulting in \fca\ observed signal.
The mass difference of \fca\ two $\cp$-even Higgs bosons can mimic a
larger width as preferred by \fca\ ATLAS data.

We have discussed several possibilities to test this scenario in the
upcoming LHC runs. These include a possible double peak structure in the
invariant $\ga\ga$ mass spectrum, due to \fca\ fact that two Higgs bosons
contribute to \fca\ signal. Also an enhanced observation of electron-jets
could be a clear signal of this scenario. 
Concerning \fca\ more ``classic'' heavy Higgs boson searches, we expect
that \fca\ relevant parameter space can be coved in \fca\ $\tau^+\tau^-$
searches, where \fca\ $750 \gev$ Higgs bosons should become detectable. On
the other hand, in our scenario {\em no} substantial decays of
the Higgs bosons at $\sim 750 \gev$ to $WW$, $ZZ$ or $Z\ga$ should be
observed.


\section*{Acknowledgements}

I thank F.~Domingo, J.S.~Kim \xyz\ K.~Rolbiecki with whom \fca\ NMSSM
results presented here have been obtained.
This work has been supported by CICYT
(grant FPA 2013-40715-P) \xyz\ in part by
Spanish MICINN's Consolider-Ingenio 2010 Program under grant 
MultiDark CSD2009-00064.


\end{document}